\documentclass[amsmath,amssymb,aps,preprintnumbers,twocolumn]{revtex4-1}
\pdfoutput=1
\usepackage{epsf}

\usepackage{epsfig,graphics}
\usepackage {graphicx}
\usepackage {epsfig}
\usepackage {subfigure}
\usepackage {tabularx} 
\usepackage{rotate}	
\usepackage{slashed}

\usepackage{amsmath}
\usepackage{amsfonts}
\usepackage{amssymb}
\usepackage{graphicx}

\newcommand{\bmat}{\left(\begin{array}}
\newcommand{\emat}{\end{array}\right)}

\def\yzero{\smash{\hbox{$y\kern-4pt\raise1pt\hbox{${}^\circ$}$}}}

\def\beq{\begin{equation}}
\def\eeq{\end{equation}}
\def\beqa{\begin{eqnarray}}
\def\eeqa{\end{eqnarray}}

\def\-{\hphantom{-}}

\def\s2{\frac{1}{\sqrt2}}

\def\beq{\begin{equation}}
\def\eeq{\end{equation}}
\def\beqa{\begin{eqnarray}}
\def\eeqa{\end{eqnarray}}

\def\IF{\relax{\rm I\kern-.18em F}}
\def\II{\relax{\rm I\kern-.18em I}}
\def\IP{\relax{\rm I\kern-.18em P}}
\def\IC{\relax\hbox{\kern.25em$\inbar\kern-.3em{\rm C}$}}
\def\IR{\relax{\rm I\kern-.18em R}}

\def\Dsl{\,\raise.15ex\hbox{/}\mkern-13.5mu D} 
\def\IZ{Z\kern-.4em  Z}

\catcode`\@=11   
\newdimen\@rotdimen
\newbox\@rotbox  

\def\@vspec#1{\special{ps:#1}}
\def\@rotstart#1{\@vspec{gsave currentpoint currentpoint translate
   #1 neg exch neg exch translate}}
\def\@rotfinish{\@vspec{currentpoint grestore moveto}}
\def\@rotr#1{\@rotdimen=\ht#1\advance\@rotdimen by\dp#1%
   \hbox to\@rotdimen{\hskip\ht#1\vbox to\wd#1{\@rotstart{90 rotate}%
   \box#1\vss}\hss}\@rotfinish}
\def\@rotl#1{\@rotdimen=\ht#1\advance\@rotdimen by\dp#1%
   \hbox to\@rotdimen{\vbox to\wd#1{\vskip\wd#1\@rotstart{270 rotate}%
   \box#1\vss}\hss}\@rotfinish}%
\def\@rotu#1{\@rotdimen=\ht#1\advance\@rotdimen by\dp#1%
   \hbox to\wd#1{\hskip\wd#1\vbox to\@rotdimen{\vskip\@rotdimen
   \@rotstart{-1 dup scale}\box#1\vss}\hss}\@rotfinish}%
\def\@rotf#1{\hbox to\wd#1{\hskip\wd#1\@rotstart{-1 1 scale}%
   \box#1\hss}\@rotfinish}%
\def\rotate{\@ifnextchar[{\@rotate}{\@rotate[l]}}
\def\@rotate[#1]#2{\setbox\@rotbox=\hbox{#2}\@nameuse{@rot#1}\@rotbox}

\catcode`\@=12

\begin{document}

\preprint{IFT-UAM/CSIC-14-034}

\title{The Inflaton as a MSSM  Higgs\\
and Open String Modulus Monodromy  Inflation}
\author{Luis E. Ib\'a\~nez}
\author{Irene Valenzuela}
\address{Departamento de F\'{\i}sica Te\'orica 
and Instituto de F\'{\i}sica Te\'orica  UAM-CSIC,\\
Universidad Aut\'onoma de Madrid,
Cantoblanco, 28049 Madrid, Spain}
\begin{abstract}
It has been recently pointed out that the polarisation  BICEP2 results are consistent
with the identification of an inflaton mass $m_{I}\simeq 10^{13}$ GeV with the 
SUSY breaking scale in an MSSM with a fine-tuned SM Higgs. This identification 
leads to a Higgs mass $m_h \simeq  126$ GeV, consistent with LHC measurements. 
Here we propose that this naturally suggests to identify the inflaton with the heavy
MSSM  Higgs system. 
The fact that the extrapolated Higgs coupling $\lambda_{SM}\simeq 0$ at scales
below the Planck scale 
suggests the Higgs degrees of freedom could be associated to
a Wilson line or D-brane position  modulus in
string theory.  The Higgs system then  has a shift symmetry and an $N=2$ structure
which guarantees that  its potential has an approximate  quadratic chaotic  inflation form. 
These moduli in string compactifications, being compact,   allow for  trans-Planckian inflaton field range
analogous to a version of monodromy inflation.
\end{abstract}
\maketitle

{\it Introduction}.
Inflationary cosmology has achieved an impressive series of successful tests, culminating with the recent polarisation measurements 
by the BICEP2 collaboration \cite{bicep2}, showing evidence for  cosmological tensor fluctuations in the very early universe.  This was a neat prediction 
of large field inflationary models and in particular of the chaotic inflationary scenario with a simple quadratic inflation potential \cite{chaotic}.
In another crucial breakthrough,  the CMS and ATLAS collaborations confirmed the existence of a Higgs boson with mass around
126 GeV \cite{Higgs}.  It has always been tempting to identify the inflaton with the Higgs boson since, after all, the Higgs boson is the only known
fundamental scalar which has been observed.  It was however soon found  that the form of the SM potential, with a quartic term and
a small mass parameter was not appropriate to generate successful inflation. Modifications were proposed  using non-minimal couplings
of the Higgs boson to the curvature, leading to inflationary models with an effective  Starobinsky like structure \cite{higgsflation}.
These models however are not free of problems
(see e.g. \cite{Barbon:2009ya} and references therein) and moreover the LHC measurement of the Higgs mass do not favour them. This is because, for the observed value of the
Higgs  and top quark masses, a RGE  extrapolation of the Higgs quartic coupling $\lambda_{SM}$ shows that  the SM potential becomes unstable at
energies of order $10^{11}-10^{13}$ GeV, well below the Planck scale \cite{elias}.
Furthermore such models predict very small tensor fluctuations,
in apparent contradiction with BICEP2 results. Small field inflation with the MSSM Higgs sector, leading to small $r$ has also
been considered in \cite{mazumdar}.

 The fact that the SM potential becomes unstable at a scale $10^{10}-10^{13}$ GeV suggests that at those scales some new physics sets on 
 stabilising the potential.  In \cite{hebecker1,imrv} it was suggested that that scale could correspond to the SUSY breaking scale and in \cite{iv} it was 
 found that indeed that assumption is consistent with the observed Higgs mass (see also \cite{hebecker2,otrosinter}).  If this is the case, the role  of SUSY would
 not be stabilising the Higgs mass, which would have to be fine-tuned \cite{hn},  but rather to stabilise the potential. From the point of view of string theory,
 the existence of SUSY at some scale, not necessarily the TeV scale, is strongly motivated, since it is a built-in symmetry of the theory and
 provides stability for the abundant scalars appearing in string compactifications. Furthermore, the fine-tuning of the Higgs mass could be 
 motivated from the point of view of the string landscape.
 
In ref \cite{iv2} we proposed that the 
polarisation  BICEP2 results are consistent
with the identification of an inflaton mass $m_{I}\simeq 10^{13}$ GeV with the 
SUSY breaking scale in an MSSM with a fine-tuned SM Higgs.  We showed how this identification led
to results for the Higgs mass consistent with the LHC results.
 Here we show that, if indeed a MSSM-like structure is realised at an intermediate scale $M_{ss}\simeq 10^{13}$ GeV,  the 
 MSSM Higgses $h,H$  can give rise to inflation.  We show that a quadratic mass term leading to a 
 chaotic inflation naturally appears if an appropriate symmetry structure is present in the Higgs sector. 
 We also show that these kind of symmetries appear in string compactifications and higher dimensional models.
 Large trans-Planckian inflaton range  may appear in a  way analogous to that  of monodromy inflation.
 
 {\it The intermediate scale MSSM and inflation}.
The minimal Higgs system in the MSSM has two EW doublets $H_u$ and $H_d$. The scalar potential for the neutral scalars  is given by a  
D-term   and general SUSY-breaking soft terms, with a general structure
 \beqa
V_{Higgs}&=& m_u^2|H_u|^2+m_d^2|H_d|^2 +(BH_uH_d+h.c.)+\nonumber \\ 
 &&\frac {g^2+g_1^3}{8}\left(|H_u|^2-|H_d|^2\right)^2 
 \label{MSSM}
\eeqa
where $m_{u,d}^2$ includes both the soft masses and a possible contribution of a SUSY $\mu$-term. 
All of them will be of order  $10^{13}$ GeV, and a massless  SM Higgs doublet would result
from a delicate fine tuning of the mass parameters   \cite{hebecker1,imrv,iv}.  
Here we will concentrate in the two complex neutral scalars.
Let us define the two eigenvalues of the mass matrix as
\beq
h=sin\beta H_u-cos\beta H_d^* \ ;\ 
H=cos\beta H_u+sin\beta H_d^*
\eeq
with respective masses $m_-^2$ and $m_+^2$ given by
\beq
m_\pm^2=\frac {1}{2} \left(m_u^2+m_d^2\pm\sqrt{(m_u^2-m_d^2)^2+4|B|^2}\right)
\eeq
Note that a zero eigenvalue, corresponding to $m_h=m_-=0$ appears when 
$|B|^2=m_u^2m_d^2$, yielding an (aproximately)  massless  Higgs $h$.
This we want to happen {\it at the SUSY-breaking scale $M_{ss}$}.
Note however that, running-up in energies to the GUT scale $m_-^2$ will be positive,
and both $m_\pm^2$ will be not vanishing at the GUT scale.
Without loss of generality we can take the neutral vevs of  $h,H$ real. 
We then get
\beqa
V_{Higgs}&=&m_-^2h^2+m_+^2H^2+\frac {g^2+g_1^2}{8}
(cos2\beta (H^2-h^2)  \nonumber \\
&+&2hH\ sin2\beta)^2 \ .
\eeqa
Note that, close below  the $M_{ss}$ scale  where $m_+\gg m_-$,   one recovers a SM Higgs potential with
\beq
V_{SM}=m_-^2h^2 + \frac {g^2+g_1^2}{8}cos^22\beta 
 |h|^4  \ .
\eeq
As we said, at  such high scales we know that the Higgs self-coupling
$\lambda\simeq 0$, which in the present context implies that $cos^22\beta\simeq 0$ at $M_{ss}$, recovering the results 
in \cite{hebecker1,imrv,iv,hebecker2}.  At scales $\gtrsim M_{ss}$ the scalar potential is then  given by
\beq
V_{Higgs}=m_-^2h^2+m_+^2H^2+\frac {g^2+g_1^2}{2}\left(
h^2H^2\right) \ .
\eeq
with $m_-^2 \lesssim m_+^2 $ and $cos^22\beta\simeq 0$, as suggested by the low-energy Higgs mass results.  
Note that along the direction $h=0$, corresponding to a very small vev for the SM Higgs,
the other MSSM Higgs scalar $H$ has a chaotic inflation scalar potential. The inflaton/Higgs potential starts with very large H vev
as in conventional chaotic inflation. The rest of the MSSM scalar Higgsses are heavy and are part of
$N=1$ massive vector multiplets.
As the inflaton goes down eventually $H$ finds a minimum at 
$H=0$, forced by the large mass $m_+$ term present at $M_{ss}$, and oscillate, reheating the universe.
Note that the reheating proceeds dominantly through SM particles.
The SM Higgs $h$  has a fine-tuned mass $m_-^2$ which is (approximately) zero around $M_{ss}$ (although 
is positive and of order $M_{ss}$ at larger scales, due to RGE running).

{\it Mass scales  and string theory}.
For this  system to work we have to check for the stability of the 
Higgs/inflaton  potential. Furthermore we know that slow-roll and large 
tensor perturbations suggest the inflaton field  should  ride along trans-Planckian 
regions. Finally, we would like to know what is the origin of the mass scale 
$m\simeq 10^{13}$ GeV which fixes  both the SUSY breaking scale and
the inflaton mass. To answer all these questions we need an UV completion 
of the theory which in what follows we assume to be string theory.

First let us discuss the origin of the SUSY breaking scale. The natural option 
in string theory is to consider the effect of antisymmetric closed string fluxes,
which for an arbitrary choice lead to SUSY breaking masses. These are 
particularly well understood in the case of Type IIB orientifold compactifications.
Consider for example a D7-brane wrapping a 4-cycle $\Sigma$ in a  compact Calabi-Yau(CY) space.
The DBI action for the brane has the general form \cite{BOOK}
\beq
S\ =\ -\ \frac {1}{g_s(\alpha')^4(2\pi)^7}\int_\Sigma d^8x \sqrt{-det(P[G+B]-2\pi\alpha'F)} \ ,
\label{DBI}
\eeq
where $g_s$ is the string coupling, $\alpha '$ the (inverse) string tension,  and $F$ is the
Yang-Mills field strength. $P[G]$ is the induced
metric and $P[B]$ the pull-back of the antisymmetric $B_{ij}$  NS field. One can integrate
locally for the $B$-field in terms of its field strength 3-form $H_3$, $B_{ij}=H_{ijk}z^k$,
with $z_k$ a coordinate in the CY transverse to $\Sigma$. This coordinate is parametrized
by the vev of a scalar $\Phi$, so that one has $B\simeq H_3\Phi$. For non-vanishing fluxes $<H_3>\not=0$,
the expansion of the action for diluted fluxes induces a mass term of the form
$<H_3^2>|\Phi |^2$. The fluxes are Dirac-quantized as $\int_{\gamma}H_3=(2\pi)^2\alpha' n_{\gamma}$,
with $\gamma$ a 3-cycle in the CY and $n_{\gamma}$ an integer. Thus  for an
isotropic compactification one expects
$H_3\simeq \alpha'/R^3$, with $R^6$ the CY volume and hence one gets
\beq m_{\Phi}^2\simeq     {H_3^2}\simeq \frac {(\alpha') 2}{R^6}
\simeq 
\frac {M_s^4}{M_p^2} \ ,
\eeq
where 
$M_s^2=(\alpha')^{-1}$, the string scale.  Taking the string scale to be 
of order the unification scale $M_s\simeq 10^{16}$ GeV, one obtains soft terms 
of the required size, $m_\Phi\simeq 10^{13}$ GeV. So the generic presence of 
antisymmetric fluxes in string theory would provide for an explanation 
for a SUSY breaking scale of that size.  Note that for a given compactification 
there is a variety of fluxes (NS, RR, non-geometric,..) that may be turned on,
leading to a variety of soft terms, see e.g. \cite{BOOK,softterms,newsoftterms}. Some particular 
classes of  fluxes may 
also give rise tu supersymmetryc couplings, like a $\mu$-term. 
All in all we have a hierarchy of mass scales
\beq 
M_{ss}\simeq 10^{13} GeV < M_c, M_{s}\simeq 10^{16} GeV < M_p \ .
\label{scales}
\eeq
Here $M_c$ is the compactification scale that, e.g. in this Type IIB setting is given by
$M_c\simeq  M_s(\alpha_{GUT}/2g_s)^{1/4}$, and hence is only slightly below the
string scale. Note in particular  that using a field theory scalar potential 
above the unification scale $10^{16}$ GeV is questionable. We discuss this point below.
In addition, if the closed string moduli are also fixed by fluxes, one should include them 
in the full scalar potential. In what follows we assume that the moduli are fixed 
at a higher scale than $M_{ss}$ respecting SUSY, so that one can consistently focus on the 
inflaton/Higgs potential. One could  obtain such a separation of scales with appropriate 
flux and geometry choices. In any event that would be very model dependent  and  we leave it  for future
investigation.

{\it Trans-Planckian inflaton}.
The large tensor perturbations detected at BICEP2 suggest a trans-Planckian field range for the inflaton \cite{lyth}.
On the other hand, as we said, using a field theory potential is questionable for field vevs above the
compactification/string scales $M_c\simeq M_s\simeq 10^{16}$ GeV. A very elegant solution to this general
problem was suggested in \cite{Silverstein} (see also  \cite{Kaloper,msu}).  If the inflation is identified with an axion-like field $a$ with
a classical shift invariance $a\rightarrow a+c$,  with a non-trivial monodromy field space, large inflation values
may be achieved without trans-Planckian axion decay constants.  In other words, the inflation range 
is not directly limited by the size of the manifold. Interestingly enough, this idea also applies 
to Wilson line and D-brane position moduli in string theory, which also present shift symmetries, in a variety of cases.

This suggests to consider the Higgs sector as associated to Wilson lines  in string
compactifications, to make the large field limit consistent in this set-up. In fact, 
as emphasised in \cite{hebecker2}, it is an intriguing fact that the apparent 
vanishing of the Higgs self-coupling at scales above $10^{11}$ GeV may be understood
in terms of a shift  invariance
\beq
H_{u,d}\rightarrow H_{u,d} + i\sigma \  \ 
\eeq
in the quadratic potential
(here $\sigma$  is real).  Indeed, the quadratic potential is only invariant if
$cos^22\beta=0$. 
Under this shift the $h$ and $H$ fields  transform as
\beq 
h\rightarrow h+i\sigma (sin\beta + cos\beta ) \ ,\  H\rightarrow H-i\sigma (sin\beta - cos\beta )  \ .
\eeq
Then for $tan\beta=1$,  $h\rightarrow h+i\sqrt{2}\sigma$ and $H\rightarrow H$,  
and a shift symmetry appears for the $h$ field.
The field $h$ in this limit is massless, 
$m_-=0$ but $m_+\not=0$.   This symmetry would not be exact in the MSSM, since e.g. 
loop corrections involving Yukawa couplings affect differently the $H_u$ and the $H_d$ masses,
but still, the fact that $\lambda_{SM}\simeq 0$ at large scales may be taken as an indication of
an approximate shift invariance at some large scale.

Analogous shift symmetries are known to be present 
in certain  subsectors of string compactifications \cite{shiftsymm}.
In particular, in heterotic  $Z_{2N}$ toroidal orbifold compactifications, the untwisted charged fields 
$H_1,H_2$ 
associated to complex planes with a twist of order two have a Kahler potential 
\beq
\kappa_4K = -log((U+U^*)(T+T^*)\ -\  \alpha' (H_1+H_2^*)(H_1^*+H_2)) \nonumber ,
\eeq
where $U$ and $T$ are the complex structure and Kahler modulus of
 the $T^2$ torus associated to the mentioned complex plane. Note that
the Kahler potential is invariant under a shift symmetry $H_{1,2}\rightarrow H_{1,2}+i\sigma$.
As noted in  \cite{shiftsymm}, if SUSY-breaking is induced by the auxiliary fields of
the moduli or the dilaton (no matter what combination), the quadratic part of the
scalar potential may be written as
\beq
V\ \propto \ (H_1+H_2^*)(H_1^*+H_2) \ ,
\eeq
which is explicitly invariant under the shift symmetry, and would correspond to a mass term $m_+^2|H|^2$ and $m_-^2=0$ 
in the Higgs case.  In the heterotic case this shift symmetry is a remnant of the gauge transformations  of a 6D gauge boson,
and the matter fields correspond to a continuous Wilson-line moduli.
This shift symmetry has also been exploited in the context of models with extra dimensions under the name of 'gauge-Higgs
unification', see \cite{gaugehiggs}.
An additional important ingredient in these string theory settings is that the   $H_{1,2}$ fields  appear in a subsector of the theory
respecting $N=2$ supersymmetry, with them forming a $N=2$ hypermultiplet.   E.g. in the heterotic  orbifold examples this happens because 
the fields $H_{1,2}$ appear from a $N=2$ sector of the compactification associated to a complex plane with an order-2 twist.  
This extended supersymmetry forbids then the appearance of any dim=4 operator   F-term contribution to the scalar potential
involving just the $H_{1,2}$ fields. 

Moduli fixing and  SUSY-breaking induced by fluxes is better understood in the context of Type II orientifolds.
Shift symmetries  for D-brane positions and/or Wilson line open string moduli also
appear in Type II  string constructions, as expected from string dualities.
Indeed, by S-duality on recovers the same structure of Wilson line moduli in Type IIB
orientifolds with D9-branes. Further T-dualities yield orientifolds with matter fields 
living on D3-branes and  or D7-branes, and Wilson lines mapping to either Wilson line moduli or
D-brane position moduli.
The latter  could  perhaps be the simplest way in trying to implement the idea of monodromy  for 
a Higgs inflation.  Recently it has been shown in \cite{msu} how 
Wilson-line monodromy inflation may be quite generic in Type II orientifold models 
with $Dp$ branes wrapping $(p-3)$-cycles in a CY (see also \cite{hebecker3,otrosaxions}).  In the simplest implementation one
can summarize the idea by saying that any periodic string moduli, either an axion, D-brane position moduli or
Wilson line moduli give rise to a monodromy potential in the presence of different types of closed
string fluxes.   For a recent discussion with
the D7-brane position acting as an inflaton see \cite{hebecker3}. 
Let us give a simple MSSM-like toy model using  an example with D7-branes on $Z_N$ singularities.

Consider a set of  6 D7-branes wrapping a 4-cycle  in a CY with local geometry $({\bf C^2} \times {\bf T^2})/Z_4$ and   located on a ${\bf Z_4}$ singularity, with local coordinates
twisted  by $(z_1,z_2,z_3)\rightarrow (\alpha z_1, \alpha z_2, \alpha^2 z_3)$, with
$\alpha=exp(i2\pi/4)$.  On the open strings there is a 
 a Chan-Paton 
matrix $\gamma=diag({\bf \alpha 1_3},\alpha^2{\bf 1_2}, {\bf 1})$. 
The open string sector includes gauge bosons in the gauge group $U(3)\times U(2)\times U(1)$ with  matter fields transforming like
(see e.g.\cite{softterms})
\beq
2(3,{\overline 2})\ +\ 2({\overline 
3},1) \ +\  (1,2)+(1,{\overline 2}) \nonumber  \ .
\eeq
To get  RR-tadpole cancellation there must be additional D3-branes at   the singularities. The open D3-D7 open strings 
complete the spectrum to two generations of the SM plus extra vector-like matter which is  not relevant for the discussion.
The main point is that, associated to the 3-d complex plane which suffers  a $Z_2$ twist, there is a vector-like set
of Higgs multiplets $(1,2)+(1,{\bar 2})$ which may be identified with $H_u, H_d$.  The vev of  $(H_u+H_d^*)$ parametrizes the location of the D7-branes
in the $z_3$ coordinate. In particular one of the two $U(2)$ D7-branes may leave the singularity if combined with the $U(1)$ D7-brane,
forming a $Z_2$ symmetric set. 
The gauge symmetry is then broken as $U(2)\times U(1)\rightarrow U(1)\times U(1)$ by the vev of $(H_u+H_d^*)$ and the remaining scalars get
massive.
We will assume that this vev has a periodic behaviour around a 1-cycle in compact dimensions. E.g., one may  consider a local
structure ${\bf C^2\times T^2}$ with $z_3$ living in the 2-torus.  One can then consider the addition of
antisymmetric RR and NS IIB closed string  fluxes, as we discussed above (one has to consider also
contributions from the D7-brane Chern-Simons action).  In the presence of  ISD $G_3$ fluxes with 
non-SUSY $G_{(0,3)}$ and SUSY  $S_{{\bar 3}{\bar 3}}$ components,  soft terms are induced, as we mentioned above.
In this example one gets soft terms of the form  \cite{softterms}
\beqa
M_3&=&M_2=M= \frac {g_s^{1/2}}{3\sqrt{2}} G_{(0,3)}^* \nonumber \\
m_{H_u}^2&=&m_{H_d}^2 = |M|^2 \ , \ \mu=-\frac {g_s^{1/2}}{6\sqrt{2}}S_{{\bar 3}{\bar 3}}^* \ ,\  B=M\mu 
\nonumber
\eeqa
where $G$ and $S$ are flux densities at the singularity, $M$ is the gaugino masses, $\mu$ is a 
SUSY mass term for the Higgs and $B$ is the standard Higgs scalar bilinear term.  These mass terms provide for
an specific origin for the MSSM scalar potential of eq.(\ref{MSSM}), with the aditional ingredient that the
Higgs fields may get large field values through the monodromy induced by the fluxes, which break the shift symmetry. 
Note that fluxes may also
induce a SUSY $\mu$-term, and that at the string/unification scale  det$(m^2_{Higgs})>0$,
and there is no zero eigenvalue.
 Such zero eigenvalue, giving rise to a light SM Higgs,  may arise however at lower energies $\simeq 10^{13}$ GeV
upon RGE running \cite{hebecker1,imrv,iv}. In the scalar potential there is a exchange symmetry $H_u\leftrightarrow H_d$
which will be eventually broken by the stronger Yukawa coupling of $H_u$ to the top quark mass, driving $m_u^2 < m_d^2$ \cite{imrv,iv}.
Although this example is not fully realistic, it illustrates how the required monodromy for the Higgs fields may easily 
appear in Type IIB models with closed string fluxes.   As we already said, a final point to remark is that in these schemes the issue of
inflation potential and the fixing of the closed string moduli are necessarily interrelated. Still one can 
play with the different volumes and 1-cycle sizes so that the moduli are fixed at scales larger than the inflaton mass \cite{msu}.
Thus a fully consistent  model should also include an appropriate treatment of the moduli fixing. Furthermore, for large inflation values the behaviour of the 
scalar potential may be modified, depending on the particular geometric implementation of the monodromy, 
see \cite{Silverstein,msu}.

{\it Discussion}. 
In this note we have proposed that the SM Higgs boson and the inflaton are SUSY partners within a MSSM structure 
at a high SUSY breaking scale.
 This leads to identify  the inflaton with the heavy scalar Higgs field $H$  which is present in the MSSM in addition to
 the standard model Higgs $h$. For this to be the case the SUSY-breaking scale should coincide with the inflaton
mass $m\simeq 10^{13}$ GeV, as recently suggested in \cite{iv2}.   Within the context of string theory, such a scale
naturally appears since for  a string scale $\simeq 10^{16}$ GeV, flux-induced soft terms are of order
$M_{ss}\simeq M_s^2/M_p\simeq 10^{13}$ GeV.  The induced mass terms give rise to a chaotic inflationary model.
The fact that the Higgs self-coupling $\lambda_{SM}$ seems
to vanish at an intermediate scale $\simeq 10^{11}-10^{13}$ GeV, suggests the existence of an approximate shift symmetry
in the MSSM Higgs system. Such type of symmetries are characteristic of  open string moduli in Type II string compactifications. 
It has been recently realised \cite{msu} that
open string  moduli, in the presence of appropriate closed string fluxes, 
naturally give rise to a simple version of
monodromy inflation (see also \cite{hebecker3}).
Large inflaton field values, as required by large field inflation and large
tensor fluctuations,  naturally appear in these schemes.
 It would be very interesting  to obtain more  complete  string compactfications in which the
Higgs doublets may be associated to a D-brane position/Wilson line moduli with non-trivial monodromy as in the toy models here suggested.
Work along these lines is in progress.

\begin{acknowledgments}

We thank   F. Marchesano,  G. Shiu  and A. Uranga for useful discussions.
This work has been supported by the ERC Advanced Grant SPLE under contract ERC-2012-ADG-20120216-320421 and 
by  the grants FPA2012-32828, and FPA 2010-20807-C02.
We also thank the  spanish MINECO {\it Centro de excelencia Severo Ochoa Program} under grant SEV-2012-0249. I.V. is supported through the FPU grant AP-2012-2690.

\end{acknowledgments}


\end{document}